\documentclass[twocolumn,showpacs,preprintnumbers,amsmath,amssymb]{revtex4}

\usepackage{graphicx}
\usepackage{dcolumn}
\usepackage{bm}

\begin{document}


\title{Repulsive and attractive Casimir interactions in liquids}

\author{Anh D. Phan}
\affiliation{University of South Florida, Tampa, USA}
\email{anhphan@mail.usf.edu}

 

\author{N. A. Viet}
\affiliation{Institute of Physics, Hanoi, Vietnam}%

\date{\today}

\begin{abstract}
The Casimir interactions in the solid-liquid-solid systems as a function of separation distance have been studied by the Lifshitz theory. The dielectric permittivity functions for a wide range of materials are described by Drude, Drude-Lorentz and oscillator models. We find that the Casimir forces between gold and silica or MgO materials are both the repulsive and attractive. We also find the stable forms for the systems. Our studies would provide a good guidance for the future experimental studies on the dispersion interactions.
\end{abstract}

\pacs{Valid PACS appear here}
\maketitle

\section{Introduction}

The dispersion interactions, the Casimir force, between neutral objects have brought attraction for many years. There are a lot of factors affecting on the value of force, such as geometry and material properties. Each of them gives rise to hot subjects of ongoing investigation. Some experiments have examined the influence of the dielectric properties of objects on the Casimir force \cite{1,2,3,6}. A number of settings used to study the interaction in terms of theory are ideal metals, real metals and semiconductors \cite{3,4,6}, metamaterials, and two objects placed in liquids \cite{1,2,5}. These studies have significantly advanced our understanding of the subtle effect of geometry and material on the Casimir-Lifshitz interactions, especially for designing nanodevices and nanotechnologies.

In the Lifshitz theory, the dispersion interactions primarily depend on dielectric permittivity functions of materials. Changing dielectric function alters the Casimir interactions. There are some ways to modify dielectric functions, including illuminating a light on the silicon \cite{7,8}, which make drifting carriers on semiconductor materials. In principle, there are some models to describe dielectric response functions of real materials, for example, plasma and Drude models for metals \cite{6,9,11}, Drude-Lorentz and oscillator models for liquids \cite{2,11}, oxides and others \cite{10,11,12}. Based on these models, the Casimir forces were obtained by numerical integrations and series expansion methods \cite{19}.

It has been theoretically shown that the attractive Casimir interaction always occurs between two (non-magnetic) dielectric bodies related by reflection. Therefore, the repulsive force is a striking feature creating inspiration for scientists to make accurate measurements of nano electromechanical machines where the repulsive force plays an important role and might resolve the stiction problems. The repulsive Casimir forces can be observed in systems which have the presence of liquids \cite{2}, metamaterials and metallic geometries \cite{16}. Recent experiments have pointed out that there a repulsive force exists between a gold sphere and a silica plate, separated by bromobenzen \cite{2}. As a matter of fact, the repulsive Casimir forces between solids arise when the dielectric of material surfaces 1 and 2 and an intervening liquid obey the relation $\varepsilon _1 (i\xi ) > \varepsilon _{liquid} (i\xi ) > \varepsilon _2 (i\xi )$ over a wide imaginary frequency range $\xi$.

A previous theoretical \cite{14} study has noticed that it is difficult to establish an equilibrium configuration of sytems in a vacuum medium. In the reference \cite{15}, the authors showed that they were able to form some stable configurations of Teflon-Si and Silica-Si immersed in ethanol. The equilibrium is explicitly explained by dispersion properties. In the present work, our theoretical studies have shown that the equilibra can be obtained by placing Au-MgO, Silica-MgO and Au-Silica sytems in bromobenzen.

In this paper, the Casimir-Lifshitz forces between material plate systems made in oxides and metals immersed in bromobenzen are calculated. The combination between these results and the proximity force approximation (PFA) method allow us to compute the Casimir interactions in different configurations. We find that the magnitude of the Casimir force between two dielectric bodies depends on the configuration and distance between two bodies. The shape usually used in experiments is a combination of a sphere and a plate because one can avoid the problem of alignment and easily control the distance between them. The energy interactions between a plate-plate system per unit area can be obtained by using the relationship between the Casimir energy of two plannar objects and the dispersion force of a sphere-plate system.     

The rest of the paper is organized as follows: In Sec. II the theoretical formulations of Casimir-Lifshitz force interaction are introduced. In section III, the numerical results for the Casimir force between two bodies are presented. Important conclusions and discussions are finally given in section IV.

\section{Lifshitz theory for force calculations}
For the force calculations, we used Lifshitz theory without considering effect of temperature. The separations used here were less than 1 $\mu$m, therefore thermal corrections at $T =300 K$ are not significant. As previously noted in \cite{2,3,20,21}, the Lifshitz formula at zero temperature for the Casimir force acting between between two parallel flat bodies per unit area, separated by a distance $d$ are given by
\begin{eqnarray}
F(d)&=&-\dfrac{\hbar}{2\pi^2} \int_{0}^{\infty}qk_{\perp}dk_{\perp}\int_{0}^{\infty}d\xi\nonumber \\
& &\times \left(\dfrac{r_{TM}^{(1)}r_{TM}^{(2)}}{e^{2qd}-r_{TM}^{(1)}r_{TM}^{(2)}} + \dfrac{r_{TE}^{(1)}r_{TE}^{(2)}}{e^{2qd}-r_{TE}^{(1)}r_{TE}^{(2)}}\right).
\end{eqnarray}
Here the reflection coefficients $r_{TM,TE}^{(1)}$ and $r_{TM,TE}^{(2)}$ for two independent polarizations of the electromagnetic field (transverse magnetic and transverse electric fields) are
\begin{eqnarray}
r_{TM}^{(p)}  = r_{TM}^{(p)} \left( {\xi ,k_ \bot  } \right) = \frac{{\varepsilon ^{(p)} \left( {i\xi } \right)q - \varepsilon ^{(2)} \left( {i\xi } \right)k^{(p)} }}{{\varepsilon ^{(p)} \left( {i\xi } \right)q + \varepsilon ^{(2)} \left( {i\xi } \right)k^{(p)} }},
\\
r_{TE}^{(p)}  = r_{TE}^{(p)} \left( {\xi ,k_ \bot  } \right) = \frac{{\mu ^{(2)} (i\xi )k^{(p)}  - \mu ^{(p)} (i\xi )q}}{{\mu ^{(2)} (i\xi )k^{(p)}  + \mu ^{(p)} (i\xi )q}},
\end{eqnarray}
where
\begin{eqnarray}
q = \sqrt {k_ \bot ^2  + \varepsilon ^{(2)} \left( {i\xi } \right)\mu ^{(2)} \left( {i\xi } \right)\frac{{\xi ^2 }}{{c^2 }}},
\\
k^{(p)}  = \sqrt {k_ \bot ^2  + \varepsilon ^{(p)} \left( {i\xi } \right)\mu ^{(p)} \left( {i\xi } \right)\frac{{\xi ^2 }}{{c^2 }}}.
\end{eqnarray}
in which $\varepsilon ^{(p)} \left( {i\xi } \right)$  and $\mu ^{(p)} \left( {i\xi } \right)$ are the dielectric permittivity and the magnetic permeability of the first body (p = 1) and the second body (p=3), respectively. $\varepsilon ^{(2)} \left( \omega  \right)$  and $\mu ^{(2)} \left( {i\xi } \right)$  are the dielectric function and the permeability of a liquid filled between two bodies. Here, medium ‘2’ selected is a bromobenzen so $\mu ^{(2)} \left( {i\xi } \right) = 1$. Moreover, in this paper, the non-magnetic materials used such as germanium, gold and oxides have also $\mu ^{(p)} \left( {i\xi } \right) = 1$. $k_ \bot$ magnitude of the wave vector component perpendicular on the plate, is frequency variable along the imaginary axis ($\omega = i\xi$).

We recall that Lifshitz formula, routinely used to interpret current experiments, express the Casimir force between two parallel plates as an integral over imaginary frequencies $i\xi$ of a quantity involving the dielectric permittivities of the plates $\omega  = i\xi$. It is important to note that, in principle, recourse to imaginary frequencies is not mandatory because it is possible to rewrite Lifshitz formula in a mathematically equivalent form, involving an integral over the real frequency axis. In this case, however, the integrand becomes a rapidly oscillating function of the frequency, which hampers any possibility of numerical evaluation. Another remarkable point is that occurrence of imaginary frequencies in the expression of the Casimir force is a general feature of all recent formalisms hence extending Lifshitz theory to non-planar geometries \cite{17,18}. The problem is that the electric permittivity $\varepsilon (i\xi )$ at imaginary frequencies cannot be measured directly by any experiment. The only way to determine it by means of dispersion relations, which allow the expression of $\varepsilon (i\xi )$ in terms of the observable real-frequency electric permittivity $\varepsilon (i\xi)$. In the standard works on the Casimir effect, $\varepsilon (i\xi )$ is expressed with the Kramers-Kronig relation in terms of an integral of a quantity involving the imaginary part of the electric permittivity \cite{13}
\begin{eqnarray}
\varepsilon (i\xi ) = 1 + \frac{2}{\pi }\int\limits_0^\infty  {d\omega \frac{{\omega {\mathop{\rm Im}\nolimits} \varepsilon (\omega )}}{{\omega ^2  + \xi ^2 }}},
\end{eqnarray}
where ${\mathop{\rm Im}\nolimits} \varepsilon (\omega )$ is calculated using the tabulated optical data for the complex index of refraction.

The well-known dielectric function described for gold is the Drude model \cite{13}
\begin{eqnarray}
\varepsilon (i\xi ) = 1 + \frac{{\omega _p^2 }}{{\xi (\xi+\gamma) }},
\end{eqnarray}
where $\omega _p = 9.0$ eV, $\gamma = 0.035$ eV are the plasma frequency and the relaxation parameter of Au, respectively.

The imaginary part of the resulting dielectric function at 6 and 295 K of pure MgO are shown in \cite{12}. The optical features have been fitted to a classical oscillator model using the complex dielectric function
\begin{eqnarray}
\varepsilon (\omega ) = \varepsilon _\infty   + \sum\limits_j {\frac{{\omega _{p,j}^2 }}{{\omega _{TO,j}^2  - \omega ^2  - i2\omega \gamma _i }}},
\label{eq:7}
\end{eqnarray}
where $\varepsilon _\infty$ is a high-frequency contribution, and $\omega _{TO,j},{2\gamma _i }$ and $\omega _{p,j}$ are the frequency, full width and effective plasma frequency of the $j$th vibration. The values of these parameters can be found in \cite{12}. Of course with such simple model for the permittivity of MgO, there is no need to use dispersion relations to obtain the expression of $\varepsilon (i\xi )$, for this can be simply done by the substitution $\omega  \to i\xi $ in the r.h.s of Eq.(\ref{eq:7}) \cite{24}
\begin{eqnarray}
\varepsilon (i\xi ) = \varepsilon _\infty   + \sum\limits_j {\frac{{\omega _{p,j}^2 }}{{\omega _{TO,j}^2  + \xi ^2  + 2\xi \gamma _j }}}.
\end{eqnarray}

In the case of bromobenzen and silica, it has recently been used for measurement of repulsive forces between gold and silica surfaces. Extremely weak repulsion was measured, indicating that the dielectric functions of bromobenzen and silica are very similar in magnitude. In fact, oscillator models are constructed to represent the dielectric function at imaginary frequencies. The form of the oscillator model is given by 
\begin{eqnarray}
\varepsilon (i\xi ) = 1 + \sum_{i}{\frac{{C _{i} }}{{1 + \xi^2/\omega_{i}^2 }}},
\end{eqnarray}
where the coefficients $C_{i}$ are the oscillator's strengths corresponding to (resonance) frequencies $\omega _{i}$ \cite{1,2,25}. The dielectric data was fitted in a wide frequency range \cite{2}. They are more accurate in comparison with other simple oscillator models. Moreover, many older references used limited dielectric data, so the oscillator models with second or third order may lead to the difference in Casimir force calculations. The parameters we used in the present paper for bromobenzen and also silica come from \cite{2}.

\section{Numerical results and discussions}
\begin{figure}[htp]
\includegraphics[width=8.5cm]{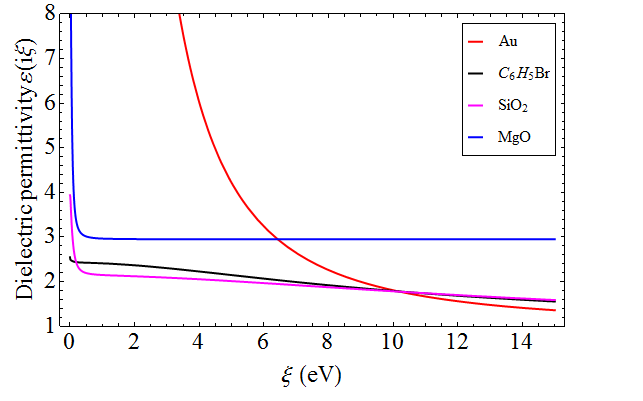}
\caption{\label{fig:1}(Color online) The dielectric function of various materials plotted at imaginary frequencies $\xi$.}
\end{figure}

The Casimir attractive force usually occurs in experiments and theoretical calculations. When bromobenzen is filled in the gap between two bodies, the  Casimir force is attractive if the dielectric functions do not satisfy one condition $\varepsilon _1(i\xi )>\varepsilon _{liquid}(i\xi )>\varepsilon _2(i\xi )$ for all frequencies $\xi$. Therefore, by describing Fig.~\ref{fig:1} as the dielectric response function as a function of the frequency gives us some predictions of repulsive and attractive forces.

\begin{figure} [htp]
\includegraphics[width=8.2cm]{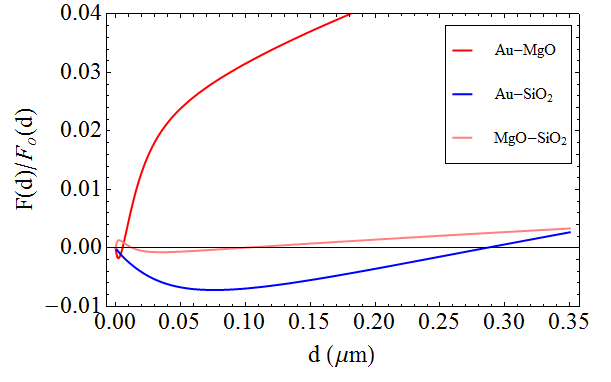}
\caption{\label{fig:2}(Color online) Relative Casimir force between two semi-infinite plates normalized by the perfect metal force $F_{o}(d)=-\pi^2\hbar c/240d^4$. The liquid used in this calculation is bromobenzen.}
\end{figure}

This graph shows that $\varepsilon_{Au}(i\xi)>\varepsilon_{MgO}(i\xi)>\varepsilon_{liquid}(i\xi)$ and $\varepsilon_{MgO}(i\xi)>\varepsilon_{Au}(i\xi)>\varepsilon_{liquid}(i\xi)$ at $\xi < 6.5$ eV, thus the interactions between Au and MgO body immersed in bromobenzen liquid and in vacuum are attractive in this range. In the range of $\xi > 6.5$ eV, $\varepsilon_{MgO}(i\xi)>\varepsilon_{liquid}(i\xi)>\varepsilon_{Au}(i\xi)$, it causes the repulsive interaction. Similarly, in the gold-bromobenzen-silica system, at extremely small frequencies , the forces are attractive. In the larger frequency region, the Casimir forces are repulsive. Besides, the similar explainations are applied to understand the interaction in the MgO-bromobenzen-Au system. The numerical calculations of the normalized Casimir force are provided in Fig.~\ref{fig:2}

In the MgO-bromobenzen-Silica system, it can be clearly seen that there are two positions in each curve where the Casimir force is equal to zero. The first points are corresponding to unstable equilibria $d_{us}^{(1)} \approx 13$ nm because the interaction force changes from the attactive force to the repulsive force, the second point $d_{s}^{(1)} \approx 110$ nm is a stable position. There is only one position in the Au-bromobenzen-Silica system and the Au-bromobenzen-MgO sytem, The interaction forces disappear at $d_{s}^{(2)} \approx 275$ nm and $d_{s}^{(3)} \approx 5.5$ nm, stable position of each sytem, respectively.

\begin{figure}[h]
\includegraphics[width=6.2cm]{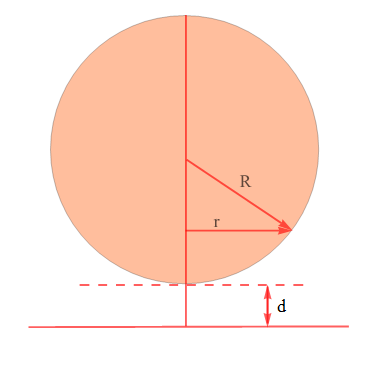}
\caption{\label{fig:3}(Color online) Schematic picture of the setting considered in our calculations. A sphere is located in bromobenzen at a distance $d$ away from a material plate.}
\end{figure}

In order to consider the Casimir interactions between a spherical body and a plate at a distance of close approach $d$ at a temperature T = 300 K, it is very useful to utilize the PFA method to calculate. Experimental results for the Casimir force in the plane-sphere geometry are usually compared with PFA-based theoretical models. The spherical surface is assumed to be nearly flat over the scale of $d$. Although the Casimir force is not additive, PFA is often expected to provide an accurate description when $R \gg d$. Here, the radius of Au sphere that is used in configurations is $R = 40$ $\mu$m in order to calculate Casimir interaction by the proximity force approximation (PFA) method because the ratio of $d$ to $R$ is small enough to PFA results becoming enormously accurate. It can be described by Fig.~\ref{fig:3}. In this approach, the surfaces of the bodies are treated as a superposition of infinitesimal parallel plates \cite{22}.

\begin{eqnarray}
F_{sp}^{PFA}(d)  = \int\limits_0^R {F_{pp} (d+R-\sqrt{R^2-r^2})} 2\pi rdr.
\label{eq:11}
\end{eqnarray}
here ${F_{pp} }$ is the Casimir force for two parallel plates of unit area.

\begin{figure}[htp]
\includegraphics[width=8.7cm]{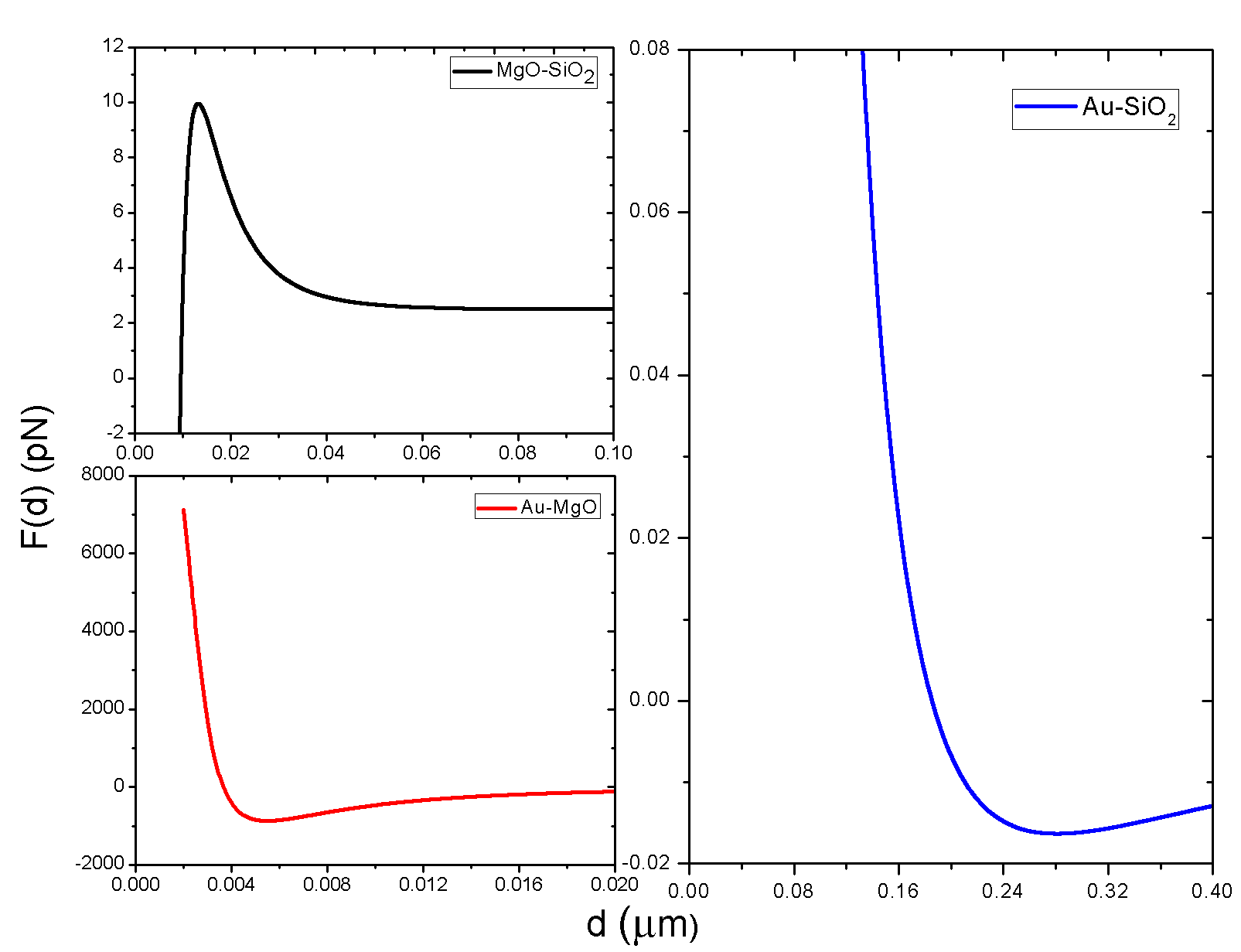}
\caption{\label{fig:4}(Color online) The Casimir forces of various sphere-bromobenzen-plate systems are estimated as a function of separation, described in the text with the spherical redius $R = 40$ $\mu$m.}
\end{figure}

When using the PFA method, one important point is that the interactions between a gold sphere or a magnesium oxide sphere and a silica plate are equal to the interactions between a magnesium oxide plate, which has the same radius, and a gold plate or a silica plate. There is no difference in calculations and results as well because the PFA method does not consider a structure of bodies when their shape is modified or is spherical or cylindrical shape. The equivalent situations occur in other materials. The resulting Casimir forces are shown in Fig.~\ref{fig:4}. 

In the reference \cite{2}, the authors experimentally measured and theoretically calculated the Casimir interaction between a gold sphere and a silica plate immersed in bromobenzen in the range from $20$ nm to $60$ nm. Our results in this range are the same for this range. But when we extend the considered range of distance, the attractive-repulsive transition occurs at approximately $190$ nm. This position makes this system stable. Another consequence of Fig.~\ref{fig:4} demonstrates that stable position of the Au-bromobenzen-MgO sytem moves to $3.5$ nm to balance between the attractive and repulsive forces. It can be explained that increasing the separation distance of infinitesimal parallel plates causes the fast reduction of the dispersion interaction. At the same minimal separation distance $d$, the attractive force acting on a sphere is less than that of a plate in the same effective area. Finally, in the system of a MgO sphere and a silica plate embeded in bromobenzen, there is only one presence of non-interaction posion at nearly $10$ nm. It is an unstable position.

\begin{figure}[htp]
\includegraphics[width=8.5cm]{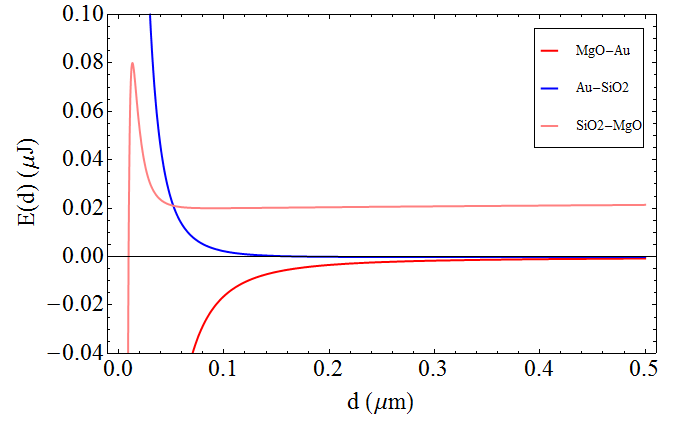}
\caption{\label{fig:5}(Color online) The Casimir energy is calculated as a function of separation for different materials .}
\end{figure}

In addition, PFA formula and Eq.(\ref{eq:11}) can allow us to estimate the Casimir energy per unit area between two plate bodies illustrated in Fig.~\ref{fig:5}. The Casimir energy is approximated by \cite{22}
\begin{eqnarray}
F_{sp}^{PFA} (d) = 2\pi R E(d),
\end{eqnarray}
where $E(d)$ is the Casimir energy per a unit area for planar bodies.

We have also applied the PFA method to calculate the Casimir force in sphere-sphere systems, we continue to calculate by the PFA method. The formula for this calculation is given
\begin{eqnarray}
& & F_{ss}^{PFA} (d) = 2\pi\int\limits_0^{R_2}rdr\times\nonumber \\
& & F_{pp}(d+R_{1}-\sqrt{R_{1}^2-r^2}+R_{2}-\sqrt{R_{2}^2-r^2}),
\label{eq:13}
\end{eqnarray}
where the radii of two spherical objects are ${R_1}$ and ${R_2}$, respectively. It is assumed that  $R_{2} < R_{1}$. In this study, we consider $R_1  = 40$ $\mu$m and the case of $R_2=R_1$, $R_1=2R_2$ and $R_1=2R_2$.

Here, having calculated $F_{ss}^{PFA}(d)$ in a sphere-sphere system using Eq.(\ref{eq:13}) and $F_{sp}^{PFA}(d)$ in a sphere-plate system using Eq.(\ref{eq:11}). These results obtained show that, when increasing $d$, the ratio $F_{ss}^{PFA} (d)/F_{sp}^{PFA} (d)$ does not depend on the distance $d$. It is a constant with its magnitude as a function of the radius of two spheres, $F_{ss}^{PFA} (d)/F_{sp}^{PFA} (d) = 1/2$ when $R_1  = R_2$ and $F_{ss}^{PFA} (d)/F_{sp}^{PFA} (d) = 1/3$ when $R_1  = 2R_2$. Generalizing this ratio, if $R_1  = nR_2$, we have $F_{ss}^{PFA} (d)/F_{sp}^{PFA} (d) = 1/(n+1)$. This character is likely to be explained by the results in \cite{22}. When the second sphere is extremely small in comparison with the first one, the interaction force goes to zero. In this case, the Lifshitz formula used to calculate the Casimir force should be transformed to the Casimir-Polder formula describing the interaction between an atom and a microscopic object. Moreover, the PFA method is not accurate because this approach is useful if the size of objects is much larger than the separation between them. On the other hand, we have 
\begin{eqnarray}
F_{ss}^{PFA} (d) = \frac{1}{n+1}F_{sp}^{PFA},
\end{eqnarray}
where $R_1  = nR_2$. If $F_{sp}^{PFA} = 0$, $F_{ss}^{PFA}$ must be zero. Therefore, the unstable and stable positions are constant and unchanged when a radius of a second sphere varies.

One demonstrated that the Casimir force between two objects embedded in liquids can be derived from the well-known Lifshitz formula at least if the object is not made of nonabsorbing materials \cite{5}. That explains why the Lifshitz expressions is used in order to calculate the Casimir force and compare with experimental data. Nothing changes in the dielectric functions of bodies immersed in liquids. On the other hand, several experiments verified that when metals are placed in liquids, there is a variation of Drude parameters in the metal \cite{23}. The discrepacy of the interaction between ``dry'' and ``wet'' can reach to 15 $\%$ in this case. But they measured the Casimir force between two metal plates and got the error. Besides, maybe the dielectric of liquids and low index materials play much more important role in Casimir force. In the reference \cite{2}, the change of Drude parameters are not taken into account but the theoretical calculations are close to the experimental data curves when we have liquids and the low index materials.
\section{Conclusions} 
In this work, we have extended the Lifshitz theory to calculate the Casimir force. Liquid, silica and magnesium oxide are represented by oscillator models. Although further studies are required to determine the repulsive Casimir force accurately, our results show that MgO and silica is a good candidate for the demonstration of quantum levitation. The contribution of bromobenzen is important because it is an important factor making the purely repulsive force or the repulsive-attractive transition. After calculating the Casimir force between two bodies per unit area and associating proximity force approximation method, it is easy to compute the interaction of different material plates with a material sphere. Based on the formula of the Casimir force between a sphere and a plate, it is convenient to estimate the free energy interaction of bodies. The result is a prediction for further experimental studies.
\begin{acknowledgments}
This work was supported by the Nafosted Grant No. 103.02.57.09. We thank Prof. Lilia M. Woods and Prof. P. J. van Zwol for helpful discussions and comments.
\end{acknowledgments}

\newpage 

\end{document}